\documentclass{article}

\usepackage[nonatbib,final]{neurips_2024}

\usepackage[numbers]{natbib}

\usepackage[utf8]{inputenc}
\usepackage[T1]{fontenc}
\usepackage{hyperref}
\usepackage{url}
\usepackage{nicefrac}
\usepackage{microtype} 
\usepackage{xcolor}

\usepackage{graphicx}
\usepackage{float}

\title{Provocation: Who benefits from ``inclusion'' in Generative AI?}

\author{
  Samantha Dalal\thanks{Equal contribution.}\\
  University of Colorado Boulder
   \\
   samantha.dalal@colorado.edu
  \And
  Siobhan Mackenzie Hall\footnotemark[1] \\
  University of Oxford \\
 siobhan.hall@nds.ox.ac.uk
   \\
  \AND
  Nari Johnson\footnotemark[1] \\
  Carnegie Mellon University\\
  narij@andrew.cmu.edu
  \\
 }

\begin{document}

\maketitle

\begin{figure}[H]
  \includegraphics[width=\textwidth]{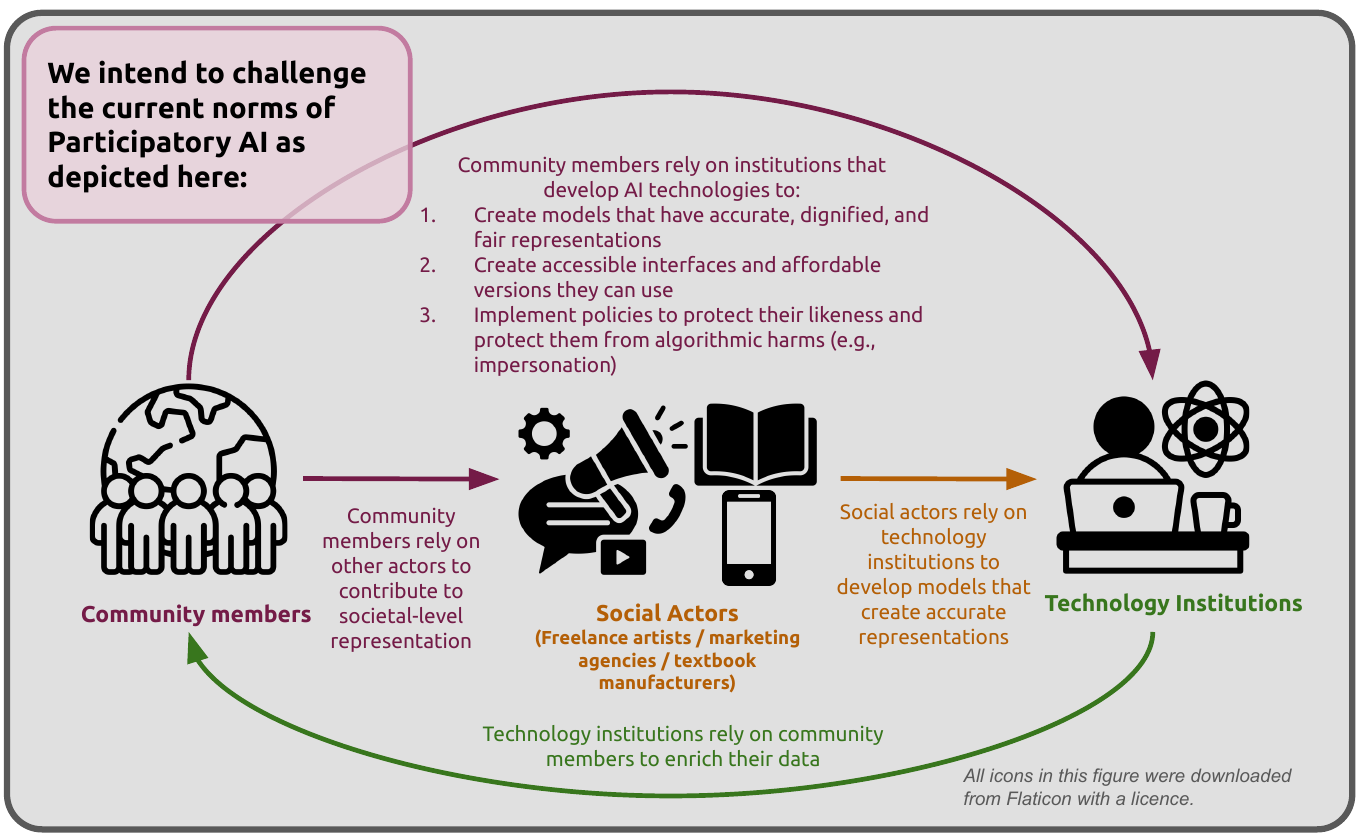}
  \caption{We map the dependencies that often exist between key stakeholder groups: (1) members of marginalized communities, (2) other social actors who use GenAI tools, and (3) institutions (such as private technology companies or academic universities) that develop GenAI models, in participatory engagements to improve how marginalized communities are represented by GenAI. In this paper, we discuss how these dependencies pose barriers for marginalized communities to realize the benefits of improved AI models.}
  \label{fig:teaser}
\end{figure}

Socially marginalized groups disproportionately experience \emph{representational harms} caused by generative AI systems  \citep{bianchi2023easily,hall2024towards,seth2024dosadatasetsocialartifacts,magomere2024eatfeedingfoundationmodels,mack2024they,katzman2023,barocas2017problem}.
For example, popular text-to-image (T2I) models have been shown to generate inaccurate, culturally misrepresentative, and insensitive depictions of racial and ethnic minorities \citep{bianchi2023easily}, people with disabilities \cite{mack2024they}, and foods from the African continent \cite{magomere2024eatfeedingfoundationmodels}.
As the AI community begins to acknowledge the limits of Internet-scraped datasets and the narrowing of values and perspectives involved in AI design procedures \citep{birhane24,birhane2022valuesencodedmachinelearning,hong2024whoswhosoutcase,devries2019does,cao-etal-2023-assessing}, there is an undeniable need to move towards including more community expertise and input. However, participation from community members runs the risk of being extractive, as many participatory engagements involve exposing participants to psychologically harmful or demeaning AI-generated content \citep{hall2024towards,magomere2024eatfeedingfoundationmodels,mack2024they, zhang2024human}.
More broadly, expertise is shifted from marginalized communities to AI model owners without commensurate structures for continued community agency and ownership over outputs~\cite{pierreGettingOurselvesTogether2021,ojewaleAIAccountabilityInfrastructure2024,birhane2022}.


In this provocation, we argue that dominant structures of community participation in AI development and evaluation are not explicit enough about the benefits and harms that members of socially marginalized groups may experience as a result of their participation~\cite{delgado2023participatoryturnaidesign}. 
Participation is increasingly motivated by trickle-down effect logics: model improvement will address stereotypes and help preserve material culture ~\cite{hall2024towards,seth2024dosadatasetsocialartifacts,magomere2024eatfeedingfoundationmodels}.
However, the potential for extractive and exploitative practices in participation is not necessarily given the same consideration.
We are concerned that the claim that community members will be better off as a result of their participation is empty, given the immensity of systemic change that is needed \textit{as well}. 
We present a speculative case study~\cite{fiesler2021innovating,Yams2021,klassen2023stoop}, based on our collective experiences doing community-engaged research in AI, to interrogate the promises AI developers make to members of marginalized groups, and itemize the barriers that realistically need to be overcome for the proposed benefits to marginalized communities to be realized.

\vspace{-20pt}
\paragraph{Speculative Case Study: How do community members realize benefits \& harms from improved GenAI performance?}\label{casestudy} Researchers do not yet fully understand how to leverage technical machine learning capabilities to improve the representation of marginalized communities in GenAI models~\cite{massiceti2024explaining,gandikota2023unified,naik2023social,seshadri2023bias}. 
Thus, we do not always know whether incorporating community feedback during the development process will necessarily lead to an ``improved'' AI model. Regardless, we believe it is important to investigate the key premise motivating participatory approaches to AI: \emph{(How) do improved representations in GenAI models benefit members of marginalized groups}?

To interrogate whom dominant structures of community participation in GenAI development benefit, we present a hypothetical scenario where a technology company invites Vietnamese community members to participate in the development and evaluation of a T2I system.
The case study that we present is meant to be an abstraction of higher-level themes that we observed over our experiences working as AI researchers within industry and academic contexts.
We use this grounding context to trace the flow of potential benefits and harms between different groups of stakeholders. 

\emph{Scenario:} Thuy, a cultural preservation activist in Vietnam, is invited to participate in a technology company's AI data enrichment initiative where community members label photographs of Vietnamese cultural artifacts for AI training and evaluation.
The company aims to improve the depiction of ``Majority World'' cultures \citep{alam2008majority} in T2I systems to support improved \textit{quality-of-service} ~\cite{shelby2023socio}, and Thuy is excited to partake in this effort to improve Vietnamese representation in global media.
The company believes that improved \textit{quality-of-service} can help Vietnamese users create images that accurately reflect their culture, for personal projects or educational purposes.
Other social actors, such as marketing agencies, textbook publishers, and freelance artists, can also benefit from generating inclusive and accurate media.
The technology company adopts a common structure of participation \cite{suresh2024participation} to engage with Thuy and other Vietnamese community advocates: they provide participants with one-time compensation for providing data and expertise.
The company has not explored paths for participant ownership or control over data or AI models that are created as a result of the engagement.



\textbf{(How) can community members benefit when they are end-users?}
Thuy and other members of her community may face \textit{financial barriers} in realizing the benefits of improved quality of service in T2I models. 
While Thuy is provided with one-time compensation for her participation, she does not continue to financially benefit from the future use of her data or the AI models it was used to improve. 
The company can improve its T2I offerings and monetize its competitive advantage by putting its services behind a paywall. 
Thuy's peers and other members can now use the model to generate accurate depictions of their likeness, but must navigate the company's paywall structures. Due to a lack of ownership over their data and resulting AI models, community ambassadors like Thuy are \emph{sold back models with improvements that would not have been possible without their labor}.

Beyond navigating paywalls, Thuy and other community members face additional barriers in \emph{accessing} and using the company's models.
For example, for many marginalized communities, model access can be complicated by several potential issues such as a lack of reliable Internet connectivity~\cite{magomere2024eatfeedingfoundationmodels} and inaccessible user interfaces \cite{vaisanen2024guidelines,das2024}. Thus, \textit{Thuy is unlikely to be able to realize the benefits of the model's improved performance as an end-user if she cannot reliably access and navigate its interface}.

\textbf{(How) can community members benefit when they are not end-users?}
Thuy may face \textit{socio-political barriers} in realizing indirect benefits resulting from social actors using T2I models as end-users to create images of the Vietnamese community. 
Many researchers have argued that due to the increasing prevalence of AI-generated media, GenAI systems will shape \textit{societal representation} ~\cite{mack2024they, katzman2023, gillespieGenerativeAIPolitics2024, noble2018algorithms,buolamwini2018gender} and thus precipitate change in societal attitudes towards marginalized communities. 
For example, a marketing firm may use the GenAI model to create images for an ad campaign that depicts Vietnamese people and culture, which is then seen by millions of people. 

However, media studies scholars have identified that representation in media \textit{alone} will not 
result in direct change to material circumstances for marginalized communities~\cite{gray2013subject,warner2017plastic,saha2012beards}. The political economy of the media ecosystem, including industry logics and financial incentives, dictates the kinds of media that are produced ~\cite{shaw2016queer}. Thus, Thuy is unlikely to realize the benefits of social actors using T2I models to create images of her community \textit{unless} social, political, and economic conditions all align to transform visibility into political power. 

\textbf{Harms marginalized groups can experience as a result of their participation}
Increased visibility in AI-generated media may make marginalized communities susceptible to a wide and emerging range of AI-mediated harms ~\cite{shelby2023socio,sigurgeirsson2024just}.
For example, as social actors (\emph{e.g.,} textbook companies) realize that they can use AI to generate accurate depictions of Vietnamese culture, they may no longer consult or compensate Vietnamese community members, resulting in further financial and social marginalization ~\cite{setty2024ai,agnew2024illusion,whitney2024real}.
Other actors can exploit improved representations of marginalized communities to inflict harm such as impersonation, misinformation, or the creation of violent/NSFW content ~\cite{romero2024generative, petit_limits_zerotolerance}.
Thus, \emph{members of marginalized communities rely on technology institutions to implement effective policies to protect their likeness}.
While the technology company could implement mitigation steps (\emph{e.g.,} access restrictions or usage licenses ~\cite{brown2023maori,licensingafrican}) to prevent misuse, they may find them at odds with their profit motives.

\textbf{Implications} 
While the details of this scenario were speculative, the discussed model of participatory engagement as one-time consultation illustrates the reality of how technology institutions and academic researchers often engage socially marginalized communities in AI development today \citep{delgado2023participatoryturnaidesign,suresh2024participation,young2024scale}.
Thus, we urge the broader AI community, including those who construct or participate in participatory engagements, to critically evaluate whether these dominant structures of participation 
do in fact yield their intended benefits \emph{for marginalized communities}. 
AI researchers and industry actors who are conducting participatory engagements with marginalized communities should be more transparent to participants and the community about the accessibility of benefits to participants and the contingencies upon which these benefits rely. 
In Appendix \ref{apdx:future-work}, we pose future directions and highlight promising examples towards restructuring participation beyond consultation, and towards supporting meaningful community ownership, participation, and power over AI.

\newpage
\section{Broader impact statement}\label{impact}

As discussed in our provocation, we believe that participatory engagements with socially marginalized groups are critical to the broader field of AI and machine learning. We urge researchers to ask themselves how communities whose participation we solicit can benefit from improved model performance. This critical self-reflection requires that researchers map out both the direct benefits participants could experience as end-users and the indirect benefits participants could realize as a result of other social actors using AI systems developed with community input. Understanding how such participatory engagements can be structured is of timely importance given the rapidly advancing capabilities of generative AI; new regulatory and policy requirements that require consultation with impacted groups~\cite{Shiming2024,hacker2023}; and to combat the increasing centralization of power in who has a say in AI's increasing influence on society \citep{birhane2022}.

\section{Limitations}\label{limitations}
We acknowledge the limitations of our analysis, which is centered around a speculative case study with imagined actors.
The barriers to realizing benefits from AI that we surfaced in our case study were based on this speculative context informed by our past experiences (\emph{e.g.,} the communities we are members of, or have engaged in research with before) and our positionality as AI researchers.
Future work can engage more deeply in analyzing real-world examples of participatory engagements with socially marginalized groups, and understanding how barriers participants face when realizing benefits vary across shared identities and contexts.

In this short piece, we briefly sketch the ``dominant structures'' of participation \cite{suresh2024participation} in GenAI evaluation, our concerns with these structures, and potential paths forward.
In doing so, our goal is \textit{not} to critique the premise that socially marginalized communities should be involved in AI development and evaluation; or that existing participatory efforts should not be pursued.
Rather, we remain hopeful that more deeply interrogating how participation is structured can lead to more empowering and constructive ways of engaging socially marginalized communities. 
Deeper engagement beyond what we could do within this workshop paper contribution is required to understand how structures of participation in GenAI development and evaluation can be shaped to support the equitable distribution of benefits and power among stakeholders.


\begin{ack}
We thank Michael Madaio, Calvin Liang, Michael Feffer, and the anonymous reviewers at the NeurIPS 2024 EvalEval Workshop for offering feedback on this work.
NJ acknowledges support from  the NSF (IIS2040929 and
 IIS2229881) and the Block Center for Technology and Society
 at CMU. 
  Any opinions, findings, conclusions, or recommendations
 expressed in this material are those of the authors and do not reflect
 the views of the National Science Foundation and other funding
 agencies.

\end{ack}
\newpage
{\small\bibliographystyle{unsrt}
\bibliography{refs}}

\newpage
\appendix
\section{Imagining paths forward}\label{apdx:future-work}
What sources of inspiration can researchers or facilitators of participatory AI initiatives turn to in their pursuit of more equitable engagement practices?

In this section, we share several resources and directions for paths forward. We do not aim to be comprehensive; rather, we highlight a few initiatives that propose alternatives to dominant participation structures. We loosely organize our discussion under two motivating questions. First, we look at theories from adjacent fields that, while not specifically about AI, provide valuable insights into participation and power. Second, we examine current efforts aimed at disrupting dominant structures and creating alternative modes of engagement that benefit marginalized communities.


\textbf{Are there theories from literature or other forms of community-based knowledge that can inform paths forward for AI?}
Participation can be extractive. Communities can lose control over how their data is used and shared once it is collected for model training, fail to be credited for their contributions to model development, or not be properly compensated for their knowledge. Below we list some resources from Indigenous data studies, dataset development, and critical data studies that identify how researchers can respect communities' preferences around data sharing. 
\begin{itemize}
    \item Christen \cite{christen2012does} identifies how some Indigenous communities have cultural norms for sharing certain types of data based on social relationships to the data artifact. AI researchers developing datasets of cultural artifacts with Indigenous communities can do work to first understand what cultural artifacts they are collecting that may have specific protocols for sharing. Researchers can then inform communities about the limitations of restricting access to images when developing and deploying GenAI models so communities can exert more informed consent. 
    \item Vincent et al. \cite{vincent2021data} conceptualize the power that contributors hold over models as data leverage. Contributors to datasets can exert power over model development and performance by reducing, stopping, redirecting, or manipulating their data. Data leverage makes explicit technology companies' \textit{dependence} on marginalized communities to improve their models' performance. Communities therefore have a significant amount of \textit{leverage} to share the terms of their future inclusion in AI development. AI researchers should consider explaining to contributors the leverage they hold over the model development process as they address fair compensation for dataset contributions. Doing so could provide contributors with a way to conceptualize the value of their data and allow them to more critically assess the remuneration they are being offered for participating and the terms of their participation. In addition, AI researchers could use data leverage to calculate more accurate estimates of financial remuneration to contributors: How much would they be willing to pay to avoid contributors using their leverage to disrupt their model? 
    
\end{itemize}

\textbf{Are there example community collaborations that offer alternative models on how to structure participation in AI development/evaluation?} 
Past scholarship \citep{delgado2023participatoryturnaidesign,suresh2024participation} has demonstrated how many ``participatory AI'' engagements are limited to consultation and inclusion (\emph{e.g.,} collecting data from participants to enrich models), without granting participants meaningful opportunities for \emph{ownership and control} over the resulting datasets and models. While participants may be able to give input on how they think the model should behave, ultimately, ``participants have little say regarding the model's impact in the world: whether it is developed, what other data it is trained on, what it may be used for, or if and how it should be deployed''~\cite{suresh2024participation}. We identify some resources where researchers and communities have been developing alternative models to structure more equitable community engagement in AI development that ensures that participants have a meaningful say over model development \textbf{and} deployment. 

\begin{enumerate}
    \item \emph{Alternative models of acknowledgment for participation}. Singh et al. ~\cite{singh2024aya} develop protocols for recognizing community contribution to AI development by operationalizing a broad definition of authorship for academic papers.
    Similar initiatives have also been led or adopted by other community-driven AI initiatives \cite{nekoto2020participatory,magomere2024eatfeedingfoundationmodels,romero2024cvqa}. Papers are a valuable currency for visibility and recognition in the AI/ML development space. By recognizing community members as contributors to AI/ML development in authorship, researchers can work towards more equitable sharing of benefits. 
    \item \textit{Alternative models of AI development}. In contrast to enriching technology companies' commercial foundation model offerings, some initiatives explore how to best support communities in developing their own smaller, more bespoke models, which are then owned and operated by community members. 
    Past efforts have surfaced how communities often need to overcome \emph{infrastructural barriers} such as limited available training data \cite{nigatu2024zeno}, capacity, and access to financial capital and compute \citep{tonja2024inkubalmsmalllanguagemodel}  to support creating, hosting, and maintaining their own models.
    \begin{enumerate}
        \item One prominent example is the Te Hiku Media foundation, a Mãori nonprofit, that decided to develop its own data hosting platform and transcription models for the \emph{te reo} language \cite{hao2022new}.
        \item  Researchers from DAIR \cite{hagdu2023combating} have similarly urged the research community to support local indigenous NLP organizations like Ghana NLP and Lesan AI who ``create machine translation systems for the specific communities they belong to''. 
    \end{enumerate}
\item \emph{Alternative models of dataset ownership and usage.} Many participatory engagements involve compensating community members in exchange for complete ownership over their data (to use for future AI development). In contrast, several communities that own their data have experimented with alternative models to govern who can use their datasets or models, and for what purpose. 
These usage restrictions are often specified in \emph{licenses} or other types of contractual agreements \cite{contractor2022behavioral,mcduff2024standardization}. 
\begin{enumerate}
    \item  Some licenses attempt to protect participants from AI-mediated harms by restricting how other stakeholders can use resulting datasets and models.
    The ``Licensing African Datasets'' project explores how to create licenses for African datasets that better redistribute benefits back towards African citizens and companies, with the expectation that ``users in developed nations would perhaps pay for use of the work or use the work under more restrictive terms'' \cite{licenseingafricandatasets}.
    Similarly, Te Hiku Media created a data license that ``will only grant data access to organizations that agree to respect Mãori values, stay within the bounds of consent, and pass on any benefits derived from use back to the Mãori people'' \cite{hao2022new}. 
    \item Future licenses can also explore specifying alternative compensation structures that allow communities to receive \emph{continued royalties} (beyond one-time compensation) to encourage profit-sharing as models that depict their likeness continue to be used \cite{deng2024computationalcopyrightroyaltymodel}.

\end{enumerate}
\item \emph{Supporting community-driven impact assessment, criticism, and refusal.} Many participatory engagements motivate community members to participate by lauding the benefits of improved GenAI representations. We urge those conducting such engagements to \emph{involve community members} in interrogating what barriers stand in the way of realizing these benefits, and in understanding potential algorithmic harms that may result from improved representations.
\begin{enumerate}
    \item Facilitators of such engagements should make room for outcomes where participants decide the harms outweigh the benefits \cite{baumer2021when,howell2021cracks}. For example, although queer scholars noticed that state-of-the-art AI voice cloning tools underperformed when cloning the voices of gay speakers, they ultimately decided against developing an improved AI technology due to concerns that an improved technology may be misused to surveil, misappropriate, or mock gay people \cite{sigurgeirsson2024just}. Making room for such critical engagements will require educating participants who enter into engagements with varying levels of familiarity about AI capabilities and harms. 
We believe that facilitators similarly have much to learn from the situated expertise of community members -- in fact, many scholars have argued that \emph{impacted communities themselves} are best equipped to anticipate AI harms \cite{moss2021assembling,kuo2023understanding,devrio2024building,sandvig2014auditing}. 
\item Communities should not just be relegated to red-teaming roles where their cultural expertise is used to identify AI harms, as this can be psychologically damaging \cite{zhang2024aura,hao2023cleaning} and further reify existing power distributions between AI developers and communities\footnote{See the reports put out by the Wiezenbaum Institute's Data Workers Inquiry for more: https://data-workers.org/}. Rather, more work is needed to build the infrastructures that empower community members to define and achieve algorithmic accountability and recourse on their own terms. For example, researchers can investigate how to support the translation of community-identified AI harms into implications for policy design to shape AI regulation following community needs. 

\end{enumerate}

\end{enumerate}



\end{document}